\documentclass[amssymb,prl,twocolumn,floats]{revtex4}
\usepackage{bm}
\usepackage{graphicx}

\begin{document}
\preprint{August 20, 2007}

\title{Magnetization dynamics down to zero field in dilute (Cd,Mn)Te quantum wells}

\author{M. Goryca$^{1}$}
\email[]{mgoryca@fuw.edu.pl}
\author{D. Ferrand$^{2}$ }
\author{P. Kossacki$^{1,2}$ }
\author{M. Nawrocki$^{1}$ }
\author{W. Pacuski$^{1,2}$}
\author{W.~Ma\'{s}lana$^{1,2}$}
\author{J.A. Gaj$^{1}$ }
\author{S. Tatarenko$^{2}$ }
\author{J. Cibert$^{2}$ }
\author{T. Wojtowicz$^{3}$ }
\author{G. Karczewski$^{3}$ }

\affiliation{
$^{1}$  Institute of Experimental Physics, University of Warsaw, Ho\.za 69, PL-00-681 Warszawa, Poland.\\
$^{2}$  Institut N\'eel, CNRS/UJF, BP166,  F-38042 Grenoble Cedex 9, France. \\
$^{3}$  Institute of Physics, Polish Academy of Sciences, Warsaw,
Poland }

\date{\today}

\begin{abstract}

The evolution of the magnetization in (Cd,Mn)Te quantum wells after
a short pulse of magnetic field was determined from the giant Zeeman
shift of spectroscopic lines. The dynamics in absence of magnetic field was found to be up to three orders of magnitude faster than that at 1~T. Hyperfine interaction and strain are mainly responsible for the fast decay. The influence of a hole gas is clearly visible: at zero field
anisotropic holes stabilize the system of Mn ions, while in a
magnetic field of 1~T they are known to speed up the decay by opening an
additional relaxation channel.

\end{abstract}

\pacs{71.35.Pq; 71.70.Gm; 75.50.Dd; 78.55.Et; 78.67.De; 85.75.-d}

\draft

\keywords{}




\maketitle


The quest for carrier-induced ferromagnetism, and for systems
appropriate to spin manipulation in the context of quantum
information processing, stimulates intense studies of Diluted
Magnetic Semiconductors (DMS)
\cite{Ohno98,Fernandez-Rossier07,Myers08}. In the DMS family,
(Cd,Mn)Te appears as particularly suitable for studies of low
dimensional structures, from quantum wells (QWs) exhibiting
carrier-induced ferromagnetic interactions \cite{Haury97}, to
quantum dots (QDs) containing a single magnetic impurity
\cite{Besombes04}. This flexibility is due to the isoelectronic
character of Mn in CdTe, offering the possibility of modulation
doping and electrical biasing of nanostructures
\cite{Boukari,Leger}, and to the broad range of optical methods
available, which yield direct information on the magnetic
configuration of the Mn system through the so called giant Zeeman
effect. While the static properties of these nanostructures are
essentially well understood, new questions arise about the
magnetization dynamics. In particular, understanding spin dynamics
in absence of any applied magnetic field is urgently needed.

Spin dynamics of Mn in (Cd,Mn)Te has been studied, for years, almost
exclusively in the presence of a magnetic field. Typical
measurements involve a short heat pulse used to drive the sample out
of thermal equilibrium, the evolution of the magnetization being
measured with a pick-up coil \cite{Strutz92} or extracted from the
giant Zeeman effect in photoluminescence (PL) \cite{Scherbakov99}.
Faraday rotation following the creation of electron-hole pairs by a
laser pulse reveals the transverse relaxation time $T_2$
\cite{Crooker96,Camilleri01}. $T_2$ was also deduced from the width
of the Electron Paramagnetic Resonance (EPR) line, in the range of
Mn content where it exhibits exchange narrowing.

\begin{figure}[]
\includegraphics[width=80mm]{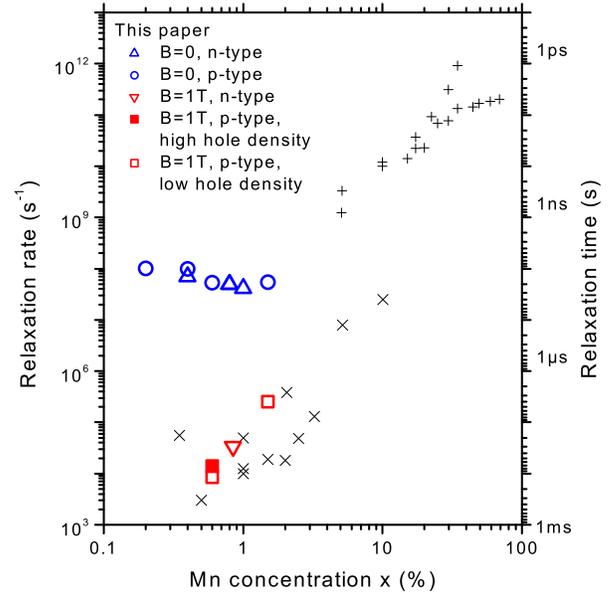}
\caption[]{spin-spin relaxation rate at 5~K (+), and spin - lattice relaxation at 1.5~K to 4.7~K ($\times$), 
and present data: fast and slow magnetization decay at 0~T and 1~T, respectively (open symbols, carrier densities in the $10^{10}~$cm$^{-2}$ range, closed symbol $3\times10^{11}~$cm$^{-2}$). Adapted from
Ref.~\cite{Dietl95}.} \label{figMncontent}
\end{figure}

The relaxation of the transverse component of magnetization is an
adiabatic process, which was ascribed to anisotropic spin-spin
interactions \cite{Larson}. It is fast, and accelerates with the Mn
content. The relaxation of the longitudinal component involves a
transfer of the Zeeman energy to the lattice (spin-lattice
relaxation). For isolated Mn$^{2+}$ ions, this is a slow process,
due to the absence of spin-orbit coupling within the orbital singlet
which forms the ground state. In DMSs, the spin-lattice relaxation
rate increases with the Mn content, as a result of a fast transfer
via the Mn spins towards "killer centers", in particular Mn
clusters \cite{Scalbert96} which statistically exist within the
random Mn distribution. For compositions where both have been
measured, the spin-spin time is faster than the spin-lattice
relaxation time, by 2 to 3 orders of magnitude (see
Fig.~\ref{figMncontent}, adapted from Ref.~\cite{Dietl95}). The
general picture that emerges, is that fast processes are expected in
a DMS with a high Mn content, while strongly diluted or isolated Mn
spins should exhibit long relaxation times. More recently
\cite{Scherbakov99}, the role of free carriers has been stressed:
they form an efficient channel to transfer the Mn Zeeman energy to
the lattice.

Experimental data at zero field are scarce. In
Ref.~\cite{Scalbert88}, the magnetization of a bulk (Cd,Mn)Te
sample, in the presence of a static field, was changed by the
collinear, AC field of a coil, and monitored by Faraday rotation. 
The spin-lattice relaxation time was clearly
identified and exhibits a dramatic decrease at low field. In
principle, this method can be used down to zero field but the
frequency response of the set-up was limited to a few microseconds,
so that the static field had to be kept above a few tenths of
tesla. By contrast, $T_2$ slowly increases when decreasing the
applied field in Zn$_{0.9}$Mn$_{0.1}$Se \cite{Crooker96} or (Cd,Mn)Te
\cite{Camilleri01}.

In this paper we describe the magnetization dynamics after a short
magnetic pulse created by a small coil, down to zero field and with
a temporal resolution of a few ns, in (Cd,Mn)Te QWs with low Mn
content, and various densities of electrons or holes.

Samples contain a single 8~nm-wide, \emph{p}-type or \emph{n}-type,
(Cd,Mn)Te QW, with 0.2\% to 1.5\% Mn, and (Cd,Mg,Zn)Te barriers. The
whole structure was grown coherently, by molecular beam epitaxy, on
a Cd$_{0.96}$Zn$_{0.04}$Te or Cd$_{0.88}$Zn$_{0.12}$Te substrate or
a CdTe buffer layer grown on GaAs. Holes were introduced into the QW
either by modulation doping with Nitrogen or by tunneling from
surface electron traps \cite{Maslana03}. Electrons were provided by modulation
doping with Iodine \cite{Wojtowicz98}. The carrier density ranged
from $1\times10^{10}$ to $3\times10^{11}$~cm$^{-2}$.

\begin{figure}[t]
\begin{center}
\includegraphics[width=85mm]{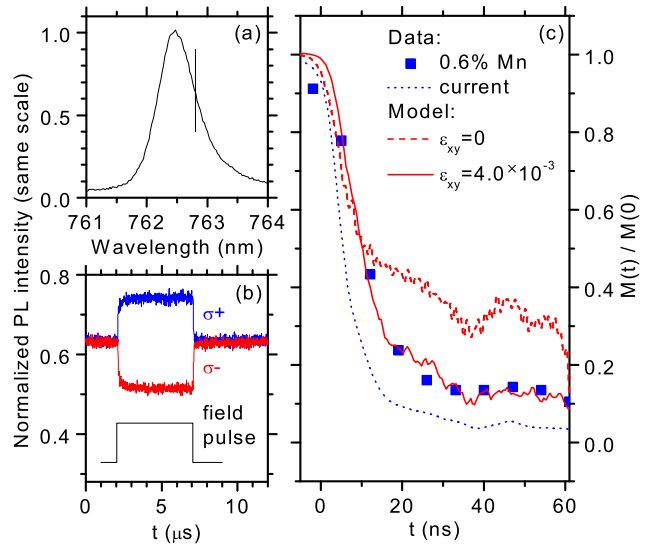}
\end{center}
\caption[]{ PL at \emph{B}=0 for a QW with 0.6\% Mn and $p\approx 10^{10}~$cm$^{-2}$. (a) Spectrum; (b) intensity \emph{vs}. time
during a field pulse, at the wavelength marked with a vertical line
in (a). (c) Magnetization dynamics (symbols), coil current (dotted line) and result of
calculations (solid and dashed lines) as described in the text.}
\label{figmethod}
\end{figure}

Pulses of magnetic field were produced with a small coil (28 turns,
internal diameter 400~$\mu$m) mounted directly on the sample
surface, fed by a coaxial transmission line
terminated with a matched resistor. A 2~A current produces
$\approx$~40~mT, as calibrated on PL spectra of samples with a high
Mn content \cite{Kossacki06}, with rise and fall times
$\approx$~7~ns (see Fig.~\ref{figmethod}). Illumination (a cw
linearly polarized diode laser) and collection of PL signal (time resolved photon counting system) were
done through the aperture of the coil, along its axis, and in some cases a
static magnetic field was added along the same axis (Faraday
configuration). 
Fig.~\ref{figmethod}a displays PL spectrum of a \emph{p}-type 
sample, obtained under typical conditions used in the
present study. It features
a single line related to the charged exciton $X^+$ (at this carrier
density, the neutral exciton is not visible). Application of a magnetic
field results in a giant Zeeman shift opposite in the two circular
polarizations, proportional to the magnetization change. If the Zeeman shift induced by the field pulse is
much smaller than the linewidth, by measuring the
intensity difference between the two polarizations, we obtain a signal proportional
to the Mn magnetization (Fig.~\ref{figmethod}b).
A higher dynamics (3 orders of
magnitude) but a slightly lower time resolution ($\approx$20~ns) was obtained using
transmission or reflectivity. Then, both the charged and neutral exciton lines
are visible, so that the field pulse induces not only a shift, but
also intensity transfers \cite{Kossacki99}. The setup for
transmission was the same as that for PL, but with a halogen lamp. In
reflectivity, we used light pulses synchronized with the field
pulse, either from a cw Ti:sapphire laser modulated to pulses
shorter than 20~ns using an acousto-optic modulator, or from a pulsed
Ti:sapphire laser. An optical bridge was used to monitor the
difference between the two circular polarizations of the reflected
light. 

Fig.~\ref{figmethod}c shows an example of the PL signal observed without applied static field. A
bi-exponential decay convoluted with the current profile describes well the magnetization decay.
The faster characteristic time is about 20~ns, and it is longer than the
experimental resolution ($\approx$7~ns). It is smaller than the values obtained in magnetic field by about
three orders of magnitude. It shows almost no dependence on the Mn content (Fig.~\ref{figMncontent}), as expected for isolated Mn ions.

Several interactions may be responsible for the fast zero-field decay observed: the hyperfine interaction with
the nuclear spin of the Mn ion itself, the superhyperfine
interaction with the nuclear spin of neighbor Cd atoms, dipolar
interaction with nuclear spins or spins of other Mn ions, the single ion anisotropy due to the
cubic crystal field and to strain in the epitaxial structure,
fluctuations in the interaction with carriers, and so on...

\begin{figure}[]
\includegraphics[width=85mm]{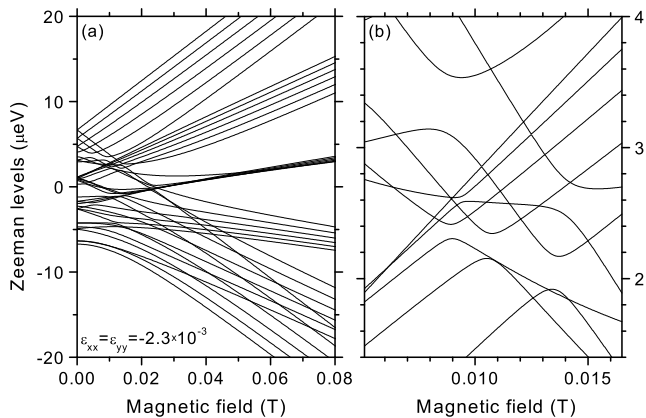}
\caption[]{Energy levels taking into account the hyperfine coupling,
cubic crystal field and mismatch strain.} \label{figenergy}
\end{figure}

It is well known, from EPR spectroscopy of Mn diluted in CdTe-
and ZnTe-MnTe superlattices \cite{Lambe,Quazzaz}, that the Mn electronic spin $\overrightarrow{S}$ interacts strongly with  its nuclear spin $\overrightarrow{I}$ (both 5/2), with a sizable crystal field, and with strain.
The Hamiltonian assessed by EPR is

\begin{eqnarray}\label{eq:eq1}
\mathcal{H}=g\mu_B\overrightarrow{B}.\overrightarrow{S}+A\overrightarrow{I}.\overrightarrow{S}
+D[S_{z}^{2}-\frac{S(S+1)}{3}] \nonumber
\\
+\frac{a}{6}[S_{x}^{4}+S_{y}^{4}+S_{z}^{4}-\frac{S(S+1)(3S^{2}+3S-1)}{5}]
\end{eqnarray}

with $A=~$680~neV for the hyperfine coupling, $a=~$320~neV for the
cubic crystal field. The dominant strain component in epitaxial layers grown pseudomorphically on substrates with 4\% and 12\% Zn is biaxial, with 
$\varepsilon_{xx}=\varepsilon_{yy}=-(c_{11}/2c_{12})\varepsilon_{zz}=-2.3\times10^{-3}$ and 
$-7\times10^{-3}$, respectively. Its effect is described by the $DS_{z}^{2}$ term, with values of $D$ deduced from Ref.~\cite{Quazzaz}. A typical energy diagram is shown in Fig.~\ref{figenergy}. 

Strong anticrossings
appear at low values of the static field (Fig.~\ref{figenergy}b): therefore sweeping down the magnetic field at the end of the pulse produces a change not only 
in the Zeeman energies but also in the state of the Mn. The evolution of the
magnetization can be viewed as a series of
Landau-Zener processes; it is easily obtained by solving numerically the
Schr\"{o}dinger equation of the density matrix, with an initial 
value corresponding to a thermalized system. As shown by the dashed line in Fig.~\ref{figmethod}c, this calculation of the evolution of the coupled Mn electronic and nuclear spins, averaged over a thermal distribution but with no additional effect of the environment, well describes the first 
fast drop in the experimental data.

The hyperfine coupling has non-zero matrix
elements only between states with $\Delta S_z=\pm1$. So several levels show no anticrossing, and the spin states remain eigenstates when varying the magnetic field. This results in the presence of a persistent magnetization which is clearly visible in Fig.~\ref{figmethod}. 

The fast decay due to the hyperfine field and the persistent magnetization are not much changed when varying slightly the value of the biaxial strain, which in our samples remains quite low. A larger value of the strain would induce a splitting of the $S=5/2$
sextuplet into three $\pm m$ doublets, see Fig.~2 in
Ref.~\cite{Quazzaz}. In particular, around zero field, the $\pm
5/2$ and $\pm3/2$ exhibit no anticrossing, only the $\pm1/2$ doublet
does. Such a strong strain exists in CdTe QD grown on ZnTe
\cite{Besombes04}, with the  $\pm1/2$ as the ground state.

A more complete decay is calculated when introducing additional mechanisms which mix states with different values of $S_z$. Within the single Mn system, this is simulated when considering strain components which are anisotropic within the QW plane. For instance, the effect of a strain $\varepsilon_{xy}=4\times10^{-3}$, calculated using Ref.~\cite{Elsa,Causa}, is shown in Fig.~\ref{figmethod}c as a solid line. It shows a good agreement with the experimental data. Such a strain in a QW could be caused by local fluctuations or dislocations, but its estimated value is surprisingly large: x-ray diffraction data gives a 4 times smaller estimate for similar structures \cite{Bodin95}. This suggests that mechanisms related to the environment of the Mn impurity should be considered.

\begin{figure}[]
\begin{center}
\includegraphics[width=85mm]{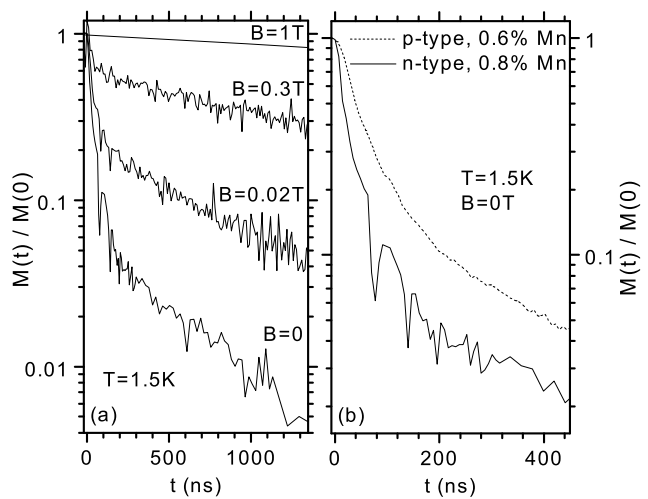}
\end{center}
\caption[]{(a) Magnetization decay in a sample with 0.8\% Mn and $n\approx 10^{10}~$cm$^{-2}$ at indicated values of the
static magnetic field. (b) Magnetization decay in \emph{n}-type and
\emph{p}-type ($p\approx 3\times 10^{11}~$cm$^{-2}$) samples at 0~T.} \label{figmagnetic}
\end{figure}

Note also that even for the best fit, the calculated evolution leaves a small but nonzero persistent
magnetization. The experimental data in
Fig.~\ref{figmethod}c and \ref{figmagnetic}a ($B = 0$) shows indeed such a long-time component, which decays with characteristic
times in sub-$\mu$s range. Upon application of an external, static magnetic field, the characteristic time 
and the amplitude of this long-time component increase. 
A precise study of this effect is beyond the scope of the present work.

It is worth to note, however, that at field of 1~T the
signal is mono-exponential, with a characteristic time in good
agreement with previous data of spin-lattice relaxation in
(Cd,Mn)Te, see Fig.~\ref{figMncontent}. In a \emph{p}-type sample,
the decay is faster for a larger hole density (which is easily controlled by additional
illumination \cite{Kossacki04}), in agreement with
Ref.~\cite{Scherbakov01}. Also the dependence of the decay at 1 T on the Mn content
(Fig.~\ref{figMncontent}) agrees with previous observations
\cite{Dietl95}. Actually, at 1~T, the pulse field merely changes the Zeeman energy
of the six sublevels of the $S=5/2$ multiplet, so that the
populations have to re-adjust through spin-lattice relaxation ~\cite{Scalbert96}.

As examplified in Fig.~\ref{figmagnetic}b, at the zero-field dynamics is systematically slower in the
presence of holes than in the presence of electrons. We ascribe this effect to the anisotropy introduced by holes. Due to
confinement, these are heavy holes with projections
$\pm3/2$ of their moment along the growth direction. Due to
the strong exchange coupling between holes and Mn spins, the
$\pm5/2$ doublet of the Mn spin (with the proper orientation of the
holes) forms the ground state, while the $\pm3/2$ and $\pm1/2$ ones are at
higher energy. A polarized hole gas of density in
the $10^{11}~$cm$^{-2}$ range induces an exchange field of the order
of 0.1~T, \emph{i.e.}, a splitting of $10~\mu$eV between the $\pm 5/2$ and
$\pm3/2$ doublets. This results in an effective anisotropy similar to that occurring 
in molecular magnets, where it gives rise to a slow dynamics
of the magnetization \cite{Bogani}. Thus the effect of
holes on the dynamics of the Mn spin at zero field
(slowing down due to an effective anisotropy) is opposite to that at high field
(additional relaxation channel). 
It was also pointed out that in a
ferromagnetic (Cd,Mn)Te QW, this coupling results in a softening
of the Mn resonance when approaching the critical temperature \cite{Kavokin,Scalbert04}, and a slow decay in the ferromagnetic phase \cite{Kossacki06}. 

To sum up, the use of a fast magnetic pulse reveals strong
peculiarities of the Mn spin dynamics in very dilute DMSs. At zero
magnetic field, a fast dynamics is observed. This dynamics is driven
by hyperfine coupling with the nuclear spin of the Mn ion; it is
highly sensitive to the presence of an anisotropy, particularly that
introduced by holes and by strain. In the present study of an
ensemble of Mn spins, with a thermal distribution of nuclear spins,
this results in a fast decay of the magnetization when the field is swept down to zero. These mechanisms
are expected to strongly influence the behavior of a single Mn spin,
with different consequences for the dephasing time observed in
time-averaged measurements and the coherence time in correlation
experiments.

\begin{acknowledgments}
We thank Denis Scalbert, Lucien Besombes, Henryk Szymczak, and Bernard Barbara for fruitful
discussions. This work was partially supported by the Polish Ministry of Science and Higher Education as research grants in years 2006-2009, Polonium and Marie-Curie programs and by the 6th Research Framework Programme of EU (contract MTKD-CT-2005-029671).
\end{acknowledgments}

\end{document}